\documentclass[twocolumn,showpacs,prc,reprint,nofootinbib,a4paper]{revtex4}  
\usepackage{epsfig}

\newcommand{\ee}{$e^{+}e^{-}$}

\newcommand{\pp}{$p+p$}

\newcommand{\agev}{$A$~GeV}
\newcommand{\gevcc}{GeV/$c^{2}$}

\newcommand{\gevc}{GeV/$c$}

\newcommand{\gev}{GeV}

\newcommand{\etal}{$et~al.$}

\begin{document}

\title{Inclusive dielectron production in proton-proton collisions at 2.2 \gev\ beam energy}

\author{
G.~Agakishiev$^{5}$, H.~Alvarez-Pol$^{15}$, A.~Balanda$^{2}$,
R.~Bassini$^{10}$,
M.~B\"{o}hmer$^{8}$, H.~Bokemeyer$^{3}$, J.~L.~Boyard$^{13}$, P.~Cabanelas$^{15}$,
S.~Chernenko$^{5}$, T.~Christ$^{8}$,
M.~Destefanis$^{9}$, F.~Dohrmann$^{4}$, A.~Dybczak$^{2}$, T.~Eberl$^{8}$,
L.~Fabbietti$^{7}$,
O.~Fateev$^{5}$, P.~Finocchiaro$^{1}$, J.~Friese$^{8}$,
I.~Fr\"{o}hlich$^{6}$, T.~Galatyuk$^{6,b}$,
J.~A.~Garz\'{o}n$^{15}$, R.~Gernh\"{a}user$^{8}$, C.~Gilardi$^{9}$,
M.~Golubeva$^{11}$, D.~Gonz\'{a}lez-D\'{\i}az$^{c}$,
F.~Guber$^{11}$, M.~Gumberidze$^{13,*}$,
T.~Hennino$^{13}$, R.~Holzmann$^{3}$, A.~Ierusalimov$^{5}$, I.~Iori$^{10,e}$,
A.~Ivashkin$^{11}$, M.~Jurkovic$^{8}$, B.~K\"{a}mpfer$^{4,d}$,
K.~Kanaki$^{4}$, T.~Karavicheva$^{11}$,
I.~Koenig$^{3}$, W.~Koenig$^{3}$, B.~W.~Kolb$^{3}$, R.~Kotte$^{4}$,
A.~Kozuch$^{2,f}$,
F.~Krizek$^{14}$, W.~K\"{u}hn$^{9}$, A.~Kugler$^{14}$, A.~Kurepin$^{11}$,
S.~Lang$^{3}$,
K.~Lapidus$^{7}$, T.~Liu$^{13}$, L.~Maier$^{8}$, J.~Markert$^{6}$,
V.~Metag$^{9}$,
B.~Michalska$^{2}$, E.~Morini\`{e}re$^{13}$, J.~Mousa$^{12}$,
C.~M\"{u}nch$^{3}$, C.~M\"{u}ntz$^{6}$, L.~Naumann$^{4}$,
J.~Otwinowski$^{2}$, Y.~C.~Pachmayer$^{6}$, V.~Pechenov$^{3}$,
O.~Pechenova$^{6}$,T.~Perez ~Cavalcanti$^{9}$, J.~Pietraszko$^{6}$, V.~Posp\'{i}sil$^{14}$,
W.~Przygoda$^{2}$, B.~Ramstein$^{13}$, A.~Reshetin$^{11}$,
M.~Roy-Stephan$^{13}$, A.~Rustamov$^{3}$,
A.~Sadovsky$^{11}$, B.~Sailer$^{8}$, P.~Salabura$^{2}$,
M.~S\'{a}nchez$^{15}$, A.~Schmah$^{a}$,
E.~Schwab$^{3}$, Yu.G.~Sobolev$^{14}$, S.~Spataro$^{g}$, B.~Spruck$^{9}$,
H.~Str\"{o}bele$^{6}$,
J.~Stroth$^{6,3}$, C.~Sturm$^{3}$, A.~Tarantola$^{6}$, K.~Teilab$^{6}$,
P.~Tlusty$^{14}$,
A.~Toia$^{9}$, M.~Traxler$^{3}$, R.~Trebacz$^{2}$, H.~Tsertos$^{12}$,
V.~Wagner$^{14}$,
M.~Wisniowski$^{2}$, T.~Wojcik$^{2}$, J.~W\"{u}stenfeld$^{4}$, S.~Yurevich$^{3}$,
Y.~Zanevsky$^{5}$, P.~Zumbruch$^{3}$}

\affiliation{
(HADES collaboration)\\\mbox{$^{1}$Istituto Nazionale di Fisica Nucleare
- Laboratori Nazionali del Sud, 95125~Catania, Italy}\\
\mbox{$^{2}$Smoluchowski Institute of Physics, Jagiellonian University of
Cracow, 30-059~Krak\'{o}w, Poland}\\
\mbox{$^{3}$GSI Helmholtzzentrum f\"{u}r Schwerionenforschung GmbH,
64291~Darmstadt, Germany}\\
\mbox{$^{4}$Institut f\"{u}r Strahlenphysik, Helmholtzzentrum
Dresden-Rossendorf, 01314~Dresden, Germany}\\
\mbox{$^{5}$Joint Institute of Nuclear Research, 141980~Dubna, Russia}\\
\mbox{$^{6}$Institut f\"{u}r Kernphysik, Goethe-Universit\"{a}t, 60438
~Frankfurt, Germany}\\
\mbox{$^{7}$Excellence Cluster 'Origin and Structure of the Universe' ,
85748~Garching, Germany}\\
\mbox{$^{8}$Physik Department E12, Technische Universit\"{a}t M\"{u}nchen,
85748~Garching, Germany}\\
\mbox{$^{9}$II.Physikalisches Institut, Justus Liebig Universit\"{a}t
Giessen, 35392~Giessen, Germany}\\
\mbox{$^{10}$Istituto Nazionale di Fisica Nucleare, Sezione di Milano,
20133~Milano, Italy}\\
\mbox{$^{11}$Institute for Nuclear Research, Russian Academy of Science,
117312~Moscow, Russia}\\
\mbox{$^{12}$Department of Physics, University of Cyprus, 1678~Nicosia,
Cyprus}\\
\mbox{$^{13}$Institut de Physique Nucl\'{e}aire (UMR 8608), CNRS/IN2P3 -
Universit\'{e} Paris Sud, F-91406~Orsay Cedex, France}\\
\mbox{$^{14}$Nuclear Physics Institute, Academy of Sciences of Czech
Republic, 25068~Rez, Czech Republic}\\
\mbox{$^{15}$Departamento de F\'{\i}sica de Part\'{\i}culas, Univ. de
Santiago de Compostela, 15706~Santiago de Compostela, Spain}\\
\\
\mbox{$^{*}$ Corresponding author. Email: sudol@ipno.in2p3.fr}\\
\mbox{$^{a}$ also at Lawrence Berkeley National Laboratory, ~Berkeley, USA}\\
\mbox{$^{b}$ also at ExtreMe Matter Institute EMMI, 64291~Darmstadt,
Germany}\\
\mbox{$^{c}$ also at Technische Universit\"{a}t Darmstadt, ~Darmstadt, Germany}\\
\mbox{$^{d}$ also at Technische Universit\"{a}t Dresden, 01062~Dresden,
Germany}\\
\mbox{$^{e}$ also at Dipartimento di Fisica, Universit\`{a} di Milano,
20133~Milano, Italy}\\
\mbox{$^{f}$ also at Panstwowa Wyzsza Szkola Zawodowa , 33-300~Nowy Sacz,
Poland}\\
\mbox{$^{g}$ also at Dipartimento di Fisica Generale and INFN,
Universit\`{a} di Torino, 10125~Torino, Italy}\\
}

\date{\today}

\begin{abstract}
Data on inclusive dielectron production are presented for the reaction \pp\ at 2.2 \gev\
measured with the High Acceptance DiElectron Spectro\-meter (HADES).  Our results supplement
data obtained earlier in this bombarding energy regime by DLS and HADES.  The comparison
with the 2.09~\gev\ DLS data is discussed.  The reconstructed \ee\ distributions 
are confronted with simulated pair cocktails, revealing an excess yield at invariant masses 
around 0.5 \gevcc.  Inclusive cross sections of neutral pion and eta production are obtained.

\end{abstract}


\maketitle

\section{Introduction}

The spectroscopy of \ee\ pairs offers a new approach to the study
of baryon resonances excited in nucleon-nucleon reactions.  Dilepton
(that is e$^{+}$e$^{-}$ or $\mu^{+}$$\mu^{-}$) observables
provide indeed information on the electromagnetic structure
of the resonances and, in the context of vector meson dominance,
their coupling to the light vector mesons \cite{faessler}.
Furthermore, dilepton spectroscopy allows to study the properties of
hadrons produced and decayed in a strongly interacting medium.
This is because leptons do not themselves interact strongly when propagating
through hadronic matter, that is, their kinematics remains basically undistorted.
For that reason they are used to probe medium modifications of hadrons
intensively searched for in photon and proton-induced reactions on nuclei
as well as in heavy-ion collisions \cite{LeupoldMetagMosel}.  Transport
models are commonly employed to describe particle production and
propagation through the medium, in particular when dealing with the
complex dynamics of nucleus-nucleus reactions \cite{hsd,urqmd,gibuu}.
The proper modeling of lepton pair production mechanisms requires a
solid understanding of the underlying elementary processes, be it in
terms of resonance excitations or in terms of a string
fragmentation picture \cite{lund}.

The HADES experiment pursues a comprehensive program of dielectron
emission studies in $N+N$ \cite{hades_nn,hades_p35p}, in $p+A$ \cite{hades_pNb},
as well as in $A+A$ collisions \cite{hades_cc2gev,hades_cc1gev,hades_arkcl}. 
Inclusive \ee\ production in $p+p$ and $p+d$ reactions had formerly been studied in
the range of 1 - 5 \gev\ by the DLS experiment at the Bevalac \cite{dls_wilson},
and more recently by HADES at 1.25~\gev\ \cite{hades_nn} and 3.5~\gev\ \cite{hades_p35p}.
In particular, the comparison of the latter data sets with various model calculations
demonstrated a need for improved theoretical descriptions.  In this paper
we supplement the available body of experimental results with data obtained
on inclusive \ee\ production in the $p+p \rightarrow p+p+e^{+}+e^{-}+X$
reaction at 2.2~\gev.  A direct comparison
with DLS data measured at 2.09~\gev\ \cite{dls_wilson} is presented.
Furthermore, through the comparison with a calculated \ee\ cocktail, we
extract the inclusive production cross sections of
$\pi^0$ and $\eta$ mesons at 2.2 \gev.  Our paper is organized as follows:
Section~2 describes the experiment and the data analysis.  In Sec.~3
\ee\ pair spectra are presented and confronted with results from DLS.
In Sec.~4 the pair spectra are compared to calculated dielectron
cocktails.  In Sec.~5 we discuss inclusive meson production cross sections
and, finally, in Sec.~6 we summarize our findings.

\section{The experiment}

The six-sector High-Acceptance DiElectron Spectrometer (HADES) operates at
the GSI Helmholtzzentrum f\"{u}r Schwerionenforschung in Darmstadt taking
beams from the heavy-ion synchrotron SIS18.
Technical aspects of the detector are described in~\cite{hades_tech}.
Its main component serving for electron and positron selection
is a hadron-blind Ring-Imaging Cherenkov detector (RICH).  Further particle identification
power is provided by the time of flight measured in a plastic scintillator
wall (TOF),  the electromagnetic shower characteristics observed in
a pre-shower detector, and the energy-loss signals from the
scintillators of the TOF wall.

In the experiment discussed here \cite{benjamin_thesis} a proton beam with a
kinetic energy of $T_p$ = 2.2~\gev\ (corresponding to a c.m. energy $\sqrt{s_{NN}} = 2.765$~\gev)
and an intensity of about $10^{7}$ particles per second impinged on a 5~cm long
liquid hydrogen cell with a total areal thickness of 0.35~g/cm$^2$.
The online event selection was done in two steps:
(1) a 1$^{st}$-level trigger (LVL1) selected events with an overall
multiplicity of at least four charged hits in the TOF wall with
additional topological conditions (two opposite sectors hit,
two hits at polar angles $<45^{\circ}$), and (2) a 2$^{nd}$-level trigger (LVL2)
required an electron or positron candidate.  This trigger scheme was
in fact optimized for measuring exclusive \ee\ production
in the $p+p \rightarrow p+p+\eta$ reaction with a subsequent $\eta$ Dalitz
decay \cite{hades_p22p}.  Note that such a trigger condition still allows
to study, albeit with a bias, inclusive \ee\ emission.  Indeed, because
of overall charge conservation in the \pp\ reaction,  the dielectron is always
accompanied by at least two more charged particles.  The resulting
trigger bias has been studied in simulations as a function of various pair
observables, in particular pair mass and pair transverse momentum, providing
a correction as well as an estimate of the resulting systematic error (of order 20\%).
For normalization purposes, \pp\ elastic scattering events were recorded concurrently
with an additional scaled-down (by a factor 32) LVL1 trigger condition requiring only
two charged hits in opposite HADES sectors.  Thus, in total $2.7
\times10^{8}$ LVL1 events were recorded, $4.1\times 10^{7}$ of which
fulfilled the LVL2 condition.

Dielectrons (that is e$^{+}$e$^{-}$ pairs) were reconstructed following the
procedures described in detail in \cite{hades_tech,hades_cc2gev}:
(1) leptons were identified based on
various detector observables, (2) an efficiency correction was applied, (3)
opposite-sign leptons were combined into pairs, (4) the background of uncorrelated
(and partially correlated) pairs representing the combinatorial background (CB)
was subtracted using the same-event geometric mean of like-sign pairs,
(5) the correction for the LVL1 trigger bias was applied, and finally
(6) the resulting inclusive \ee\ distributions were normalized to
the reconstructed yield of elastically scattered protons into
the HADES geometric acceptance (see \cite{hades_p22p} for details). 
As no dedicated start detector was present in this experimental run, the
start time for the time-of-flight measurement was reconstructed
event-by-event from the most optimal fit of different event hypotheses
to the global event data \cite{hades_p22p}.

\section{Results}

\subsection{Invariant mass spectra}

\begin{figure}[!htb]

  \mbox{\epsfig{width=0.99\linewidth, height=0.99\linewidth, figure=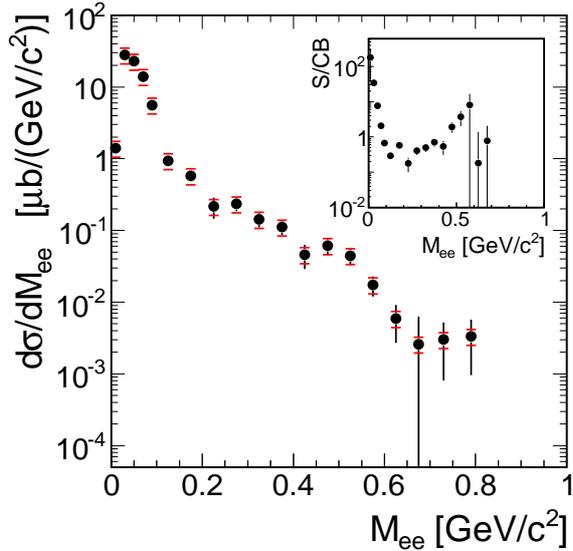}}
  \vspace*{-0.2cm}
  \caption[]{(Color online) Differential \ee\ cross section $d\sigma/dM_{ee}$
    measured in the 2.2 \gev\ \pp\ reaction within the HADES acceptance
    (including $p_e > 0.1$ \gevc\ and $\theta_{ee} > 9^{\circ}$ cuts).
    The data are efficiency corrected and CB subtracted; the insert shows
    the signal-over-CB ratio (S/CB).  Point-to-point statistical and
    systematic errors are indicated by (black) vertical bars and (red) horizontal ticks, respectively.
    }
  \vspace*{-0.2cm}
  \label{p22p_mass}
\end{figure}

\begin{figure}[!htb]

  \mbox{\epsfig{width=0.99\linewidth, height=0.99\linewidth, figure=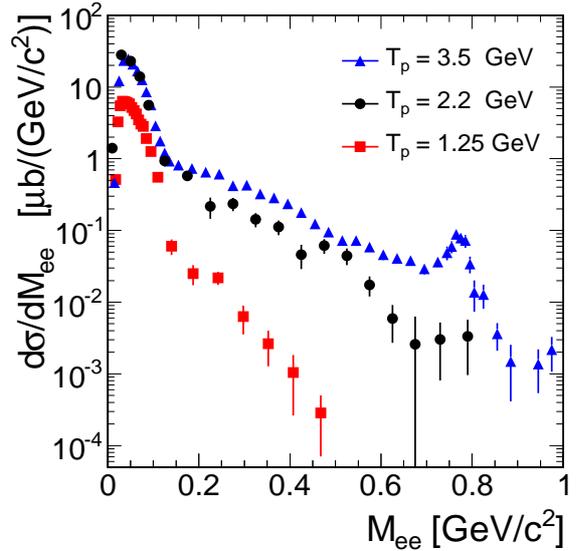}}
  \vspace*{-0.2cm}
  \caption[]{(Color online) Systematics of \ee\  differential production cross sections
    $d\sigma/dM_{ee}$ measured in \pp\ reactions at 1.25 \gev\ (squares),
    2.2 \gev\ (circles), and 3.5 \gev\ (triangles), all obtained
    within the HADES acceptance, efficiency corrected, and CB subtracted
    (including $p_e > 0.1$ \gevc\ and $\theta_{ee} > 9^{\circ}$ cuts).  
    The 1.25 \gev\ data, taken from \cite{hades_nn}, are adjusted
    for the present, more restrictive detector acceptance (i.e. stronger magnetic
    field and explicit lepton momentum cut of 0.1~\gevc); the 3.5 \gev\ data
    are taken from \cite{hades_p35p}.  Only statistical error bars are shown.
    }
  \vspace*{-0.2cm}
  \label{pp_systematics}
\end{figure}

Figure~\ref{p22p_mass} shows the differential \ee\ cross section $d\sigma/dM_{ee}$
obtained after correcting the reconstructed pair yield for efficiency,
combinatorial background, and trigger bias.  As explained above, the absolute
normalization was done using the known \pp\ elastic scattering cross section with
the help of the concurrently measured yield of elastic events \cite{hades_p22p}.
The data are presented with analysis cuts on single-lepton momentum,
$p_e > 0.1$ \gevc, and on pair opening angle, $\theta_{ee} > 9^{\circ}$.
The total number of \ee\ signal pairs contributing to this spectrum is
around 19,000 from which 2,000 pairs are located above the $\pi^0$ Dalitz
region ($M_{ee} > 0.15$ \gevcc).  To illustrate the significance of the
reconstructed dielectron signal, the signal-over-CB ratio is also shown as an
inset in Fig.~\ref{p22p_mass}.  Note that the kinematic cutoff corresponding
to the beam energy of 2.2~\gev\ is at a pair mass of 0.89~\gevcc.

The obtained cross section is combined in Fig.~\ref{pp_systematics} with
inclusive data from other HADES \pp\ runs done at 1.25 \cite{hades_nn}
and 3.5 \gev\ \cite{hades_p35p} bombarding energy, respectively.  While the 3.5~\gev\
data were recorded in the same detector acceptance as the present 2.2~\gev\ data,
the 1.25~\gev\ data were adjusted for differences in the magnetic
field strengths and analysis cuts used.  This way the three data sets are
compared within the same instrumental acceptance, revealing the strong beam energy
dependence of dielectron production, particularly at large pair masses.
Note, however, that between the three beam energies the average momentum
of the involved single lepton distributions differs significantly, resulting in
substantially different fractions of accepted pairs, particularly for low masses.

\subsection{Comparison with DLS}

\begin{figure}[!ht]

  \mbox{\epsfig{width=0.90\linewidth, figure=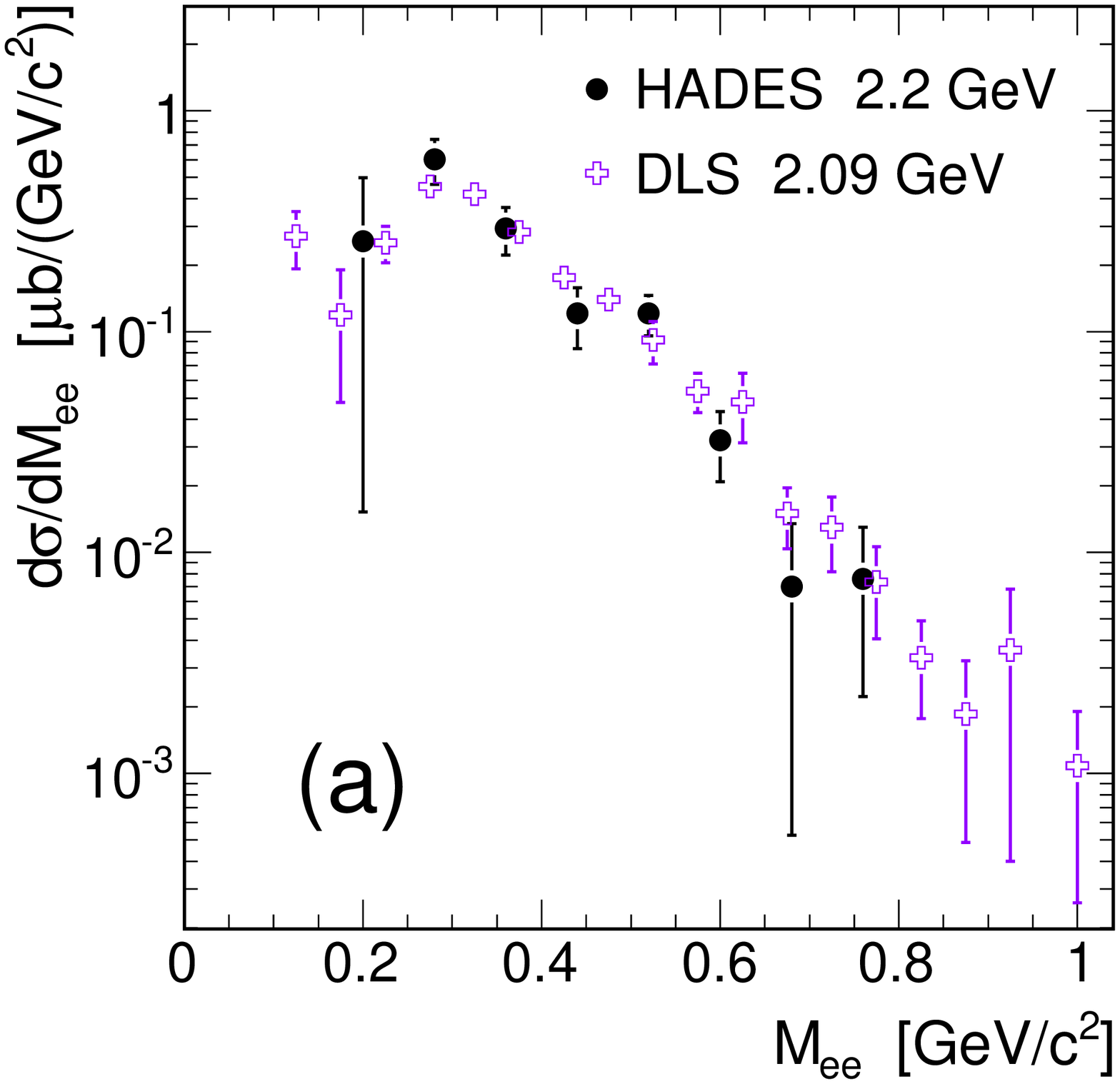}}
  \vspace*{-0.2cm}
  \mbox{\epsfig{width=0.90\linewidth, figure=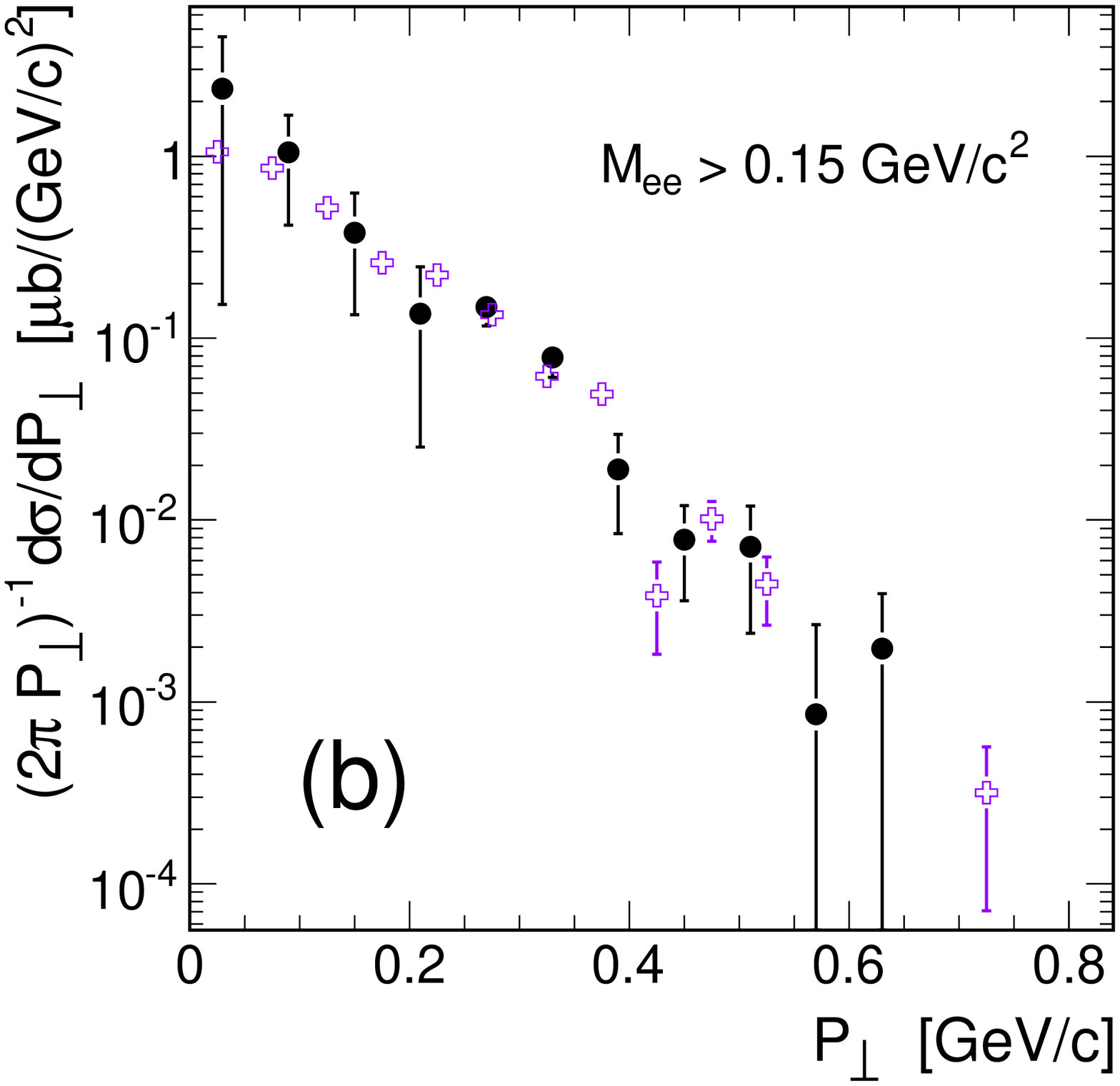}}
  \vspace*{-0.2cm}
  \caption[]{(Color online) Direct comparison within the DLS acceptance
   (see text for details) of the \ee\ cross sections
    measured by HADES in \pp\ at 2.2~\gev\ (closed circles) and by DLS at
    2.09~\gev\ (open crosses, taken from \cite{dls_wilson}).
    The pair mass distributions (a) and pair transverse-momentum
    distributions (b) are confronted.  Error bars are statistical only;
    additional systematic errors (not shown) are 23\% for DLS
    \cite{dls_wilson} and 29\% for HADES. 
    }
  \vspace*{-0.2cm}
  \label{hades_dls}
\end{figure}

As the former DLS experiment provided \ee\ data for the \pp\ reaction at
2.09~\gev\ \cite{dls_wilson}, we can make a direct comparison along the line
already used to confront the DLS and HADES C+C data obtained
at 1\agev\ \cite{hades_cc1gev,dls_porter}.
As the HADES geometrical acceptance is substantially broader than the DLS
one such a comparison can be done by projecting the reconstructed HADES
dielectron yields $d^3N/dM_{ee}dP_{\perp}dY$ through the DLS acceptance
filter \cite{dls_porter} (here $P_{\perp}$ and $Y$ are the \ee\ pair transverse
momentum and rapidity, respectively).  In fact, as DLS applied to their \pp\ data
additional cleaning cuts \cite{dls_wilson,dls_matis} -- $0.1<M_{ee}<1.25$ \gevcc,
$P_{\perp}<1.2$ \gevc, $0.5<Y<1.7$, and $\theta_{e}>21.5^{\circ}$ --
an extrapolation of the HADES yield to rapidities above 1.9,
as applied in \cite{hades_cc1gev}, is not needed here.
The result of this filtering procedure is shown in Fig.~\ref{hades_dls}(a)
for the pair mass distributions $d\sigma/dM_{ee}$ and in (b) for the pair
transverse momentum spectrum $1/(2\pi P_{\perp}) \; d\sigma/dP_{\perp}$,
the latter one with the condition $M_{ee}>0.15$ \gevcc.

It is apparent that, within statistical and systematic uncertainties, the
HADES and the DLS data are in good agreement.  This result suggests that
our result, together with the data obtained by the DLS energy scan \cite{dls_wilson},
can be used to constrain the various models aimed at describing dielectron production in
few-\gev\ elementary reactions \cite{hades_p35p}.


\section{Comparison with a simulated cocktail}

The experimental pair distributions are next compared to a calculated
\ee\ cocktail.  For this, \pp\ reactions were simulated with the event generator
Pluto \cite{pluto,pluto2} and filtered through the HADES acceptance.
The simulation included the following \ee\ pair sources:
(1) $\pi^0 \rightarrow \gamma e^+ e^-$, (2) $\eta \rightarrow \gamma e^+ e^-$,
(3) $\Delta(1232) \rightarrow N e^+ e^-$, (4) $\omega \rightarrow \pi^0 e^+ e^-$,
(5) $\omega \rightarrow e^+ e^-$, and (6) $\rho^0 \rightarrow e^+ e^-$
with dielectron branching ratios of mesons taken from \cite{pdg2010} and
that of the $\Delta(1232)$ as calculated in \cite{faessler}.
At the present bombarding energy the production of $\pi^0$ and $\eta$
mesons is known to proceed mostly via resonance excitation (e.g. $R$ = $\Delta(1232), N^*(1440),
N^*(1520), N^*(1535)$, etc.) and is dominated by one-meson and two-meson
channels \cite{hades_p22p,bystricky,moskal,cao}.  For $\rho^0$ and $\omega$ production
we have, however, assumed phase-space population in the $p p \rightarrow p p X$ reactions
($X = \rho^0$ or $\omega$) with no further attempt at a more refined
description of the high-mass region.  Note that some of the excited
resonances $R$, mostly the $\Delta$, contribute also directly to the dielectron
yield via their electromagnetic Dalitz decay $R \rightarrow N e^+ e^-$.  In our cocktail
calculation we have only taken into account the $\Delta^0$ and $\Delta^+$ contributions
following the prescription from Ref.~\cite{hades_p35p}.
The production cross sections used in the simulation were taken as
follows: 
\begin{enumerate}
\item Inclusive $\pi^0$ production -- 14~mb (adding all observed
inelastic channels contributing to $\pi^0$ production from Ref.~\cite{baldini}
gives a lower limit of about 12~mb, whereas 14~mb are needed to fully
exhaust the measured Dalitz yield).
\item Inclusive $\eta$ production -- in the range of 0.26 - 0.35~mb;
a lower limit of 0.14~mb is given by the known exclusive $\eta$
production \cite{hades_p22p,balestra}.
\item Vector meson production -- 0.01~mb exclusive $\omega$ production
\cite{cosytof} and assuming likewise for the $\rho^0$, but no $\phi$ contribution.
\item Inclusive $\Delta^{0,+}(1232)$ excitation -- in the range 10 - 21~mb.
\end{enumerate}
The extremes of the cross section range used for $\Delta$ production
correspond to the following two scenarios: (I) assumes that pion production
is mediated completely by single $\Delta$ excitation only, that is
$\sigma_{\Delta} = 3/2 \, \sigma_{\pi^0}$ (from isospin addition rules), resulting
in the upper value of 21~mb;  scenario (II) sums explicitly the $\Delta$
contributions from one-pion and two-pion production channels.  In the latter
the one-pion part of 3.6~mb is taken from a resonance model fit
to exclusive pion production data \cite{hades_p22p} whereas the two-pion part
of 6.4~mb is taken from the effective Lagrangian model of two-pion production
presented in Ref.~\cite{cao}.  Under the assumption and as suggested
indeed by various calculations \cite{hsd,urqmd,gibuu} that
dielectron production is dominated in the mass range 0.15 - 0.45~\gevcc\
by the $\Delta$ and $\eta$ contributions, the total yield measured for these
masses can be used to constrain the $\eta$ contribution for any
assumed $\Delta$ cross section.  In other words, the $\eta$ and $\Delta$
contributions are complementary.  The extracted $\eta$ cross section will
evidently have a model dependence which, however, turns out to be small.

\begin{figure}[!htb]

  \mbox{\epsfig{width=0.99\linewidth, figure=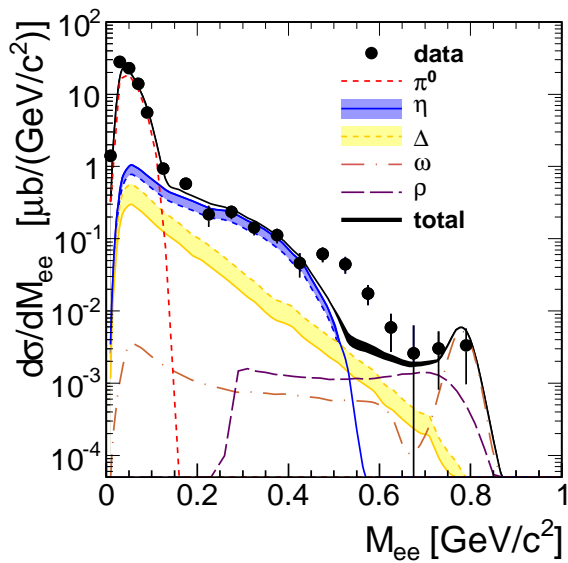}}
  \vspace*{-0.2cm}
  \caption[]{(Color online) Pair mass distribution $d\sigma/dM_{ee}$
    measured in 2.2~\gev\ \pp\ reactions (full circles) compared with
    simulated Pluto \cite{pluto,pluto2} cocktails of dielectron sources
    filtered through the HADES acceptance.  The shaded bands delimit
    the range of modeled $\Delta$ and $\eta$ contributions -- as discussed in
    the text -- with the dashed delimiters corresponding to scenario (I) and the
    solid ones to scenario (II).  Only statistical errors are shown.
    }
  \vspace*{-0.2cm}
  \label{p22p_mass_pluto}
\end{figure}

\begin{figure*}[!htb]
  \mbox{\epsfig{width=0.95\linewidth, figure=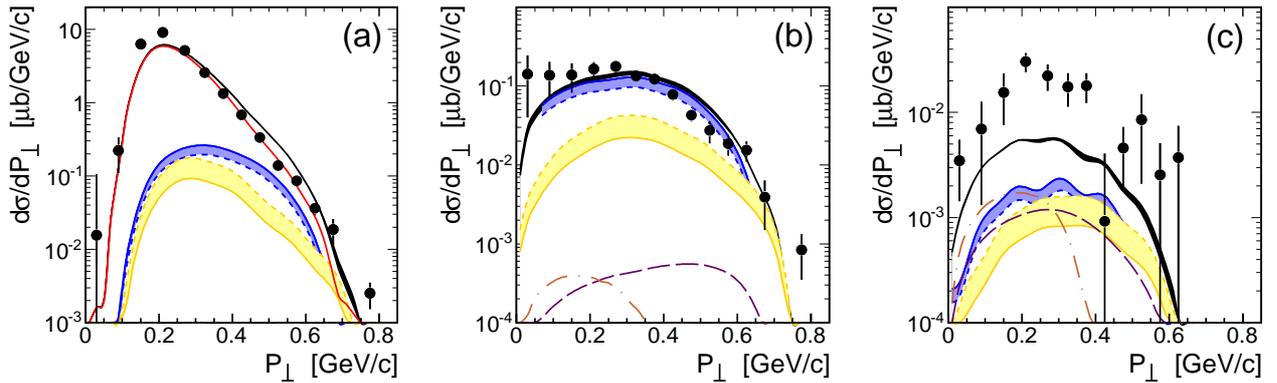}}
  \vspace*{-0.2cm}
  \caption[]{(Color online) Pair transverse momentum distributions $d\sigma/dP_{\perp}$
    measured in 2.2~\gev\ \pp\ reactions within the HADES acceptance.  Three
    mass bins are shown:
    (a) $M_{ee} < 0.15$, (b) $0.15 < M_{ee} < 0.45$, and (c)  $M_{ee} > 0.45$~\gevcc.
    The simulated Pluto cocktails are shown as well, with line
    styles as in Fig.~\ref{p22p_mass_pluto}.
    }
  \vspace*{-0.2cm}
  \label{p22p_pt_pluto}
\end{figure*}

The resulting \ee\ cocktails, filtered with the HADES acceptance, are overlayed
in Figs.~\ref{p22p_mass_pluto} and \ref{p22p_pt_pluto} with the data.  Up to
masses of $\simeq0.45$~\gevcc\ the agreement is good in both observables,
$M_{ee}$ and $P_{\perp}$,  although at higher masses the measured yield is not
matched.  Integrating up to 0.15~\gevcc\ the low-mass region, dominated
evidently by the $\pi^0$ Dalitz contribution, and correcting for the
detector acceptance, allows to fix the inclusive $\pi^0$ production cross section
at $\sigma_{\pi^0} = 14 \pm 3.5$~mb.  The quoted 25\% error is determined mostly
by systematic effects (normalization, trigger bias correction, acceptance correction).
In a similar way, the integrated yield from the mass region of 0.15 - 0.45~\gevcc\
has been used to extract the cross section of inclusive $\eta$ production after
correcting, as stated above, for the $\Delta$ contribution.
The range of assumed $\Delta$ cross sections used in the simulation leads
consequently to a corresponding range of $\eta$ production cross sections,
indicated by the hatched bands shown in the two figures.  In scenario (I),
where all $\pi^0$ production goes through $\Delta$ excitation and decay,
$\sigma_{\Delta}$ = 21~mb and $\sigma_{\eta} = 0.26$~mb.
In scenario (II), based on the model of \cite{cao}, various nucleon resonances
contribute to pion production, such that $\sigma_{\Delta} = 10$~mb
and $\sigma_{\eta} = 0.35$~mb.  We feel that this model dependence is
relatively small and propose to use the average of the two
results, namely $\sigma_{\eta} = 0.31 \pm 0.08 \pm 0.05$~mb, where
the first error is again mostly ruled by systematic effects
(normalization, trigger bias and acceptance corrections) while the second one
covers the model dependence.

In the pair mass range 0.45 - 0.60~\gevcc\ our simulation underestimates grossly the
observed yield.  This is also visible in the comparison of transverse momentum
distributions shown on Fig.~\ref{p22p_pt_pluto}. 
Clearly additional dielectron sources are needed, among
which one has to consider the decays of $N^*$ resonances, e.g. the
$N^*(1520)$ and $N^*(1720)$ known to couple strongly to the $\rho$,
as well as a possible general enhancement due to vector meson dominance
form factors of the nucleon resonances \cite{faessler}.

\section{Inclusive meson production}

\begin{figure}[!htb]

  \mbox{\epsfig{width=0.95\linewidth, height=0.90\linewidth, figure=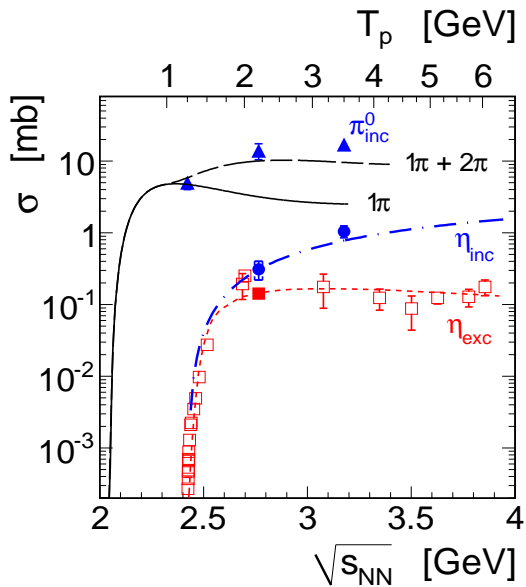}}
  \vspace*{-0.2cm}
  \caption[]{(Color online) $\pi^0$ and $\eta$ production cross sections
    in \pp\ reactions as a function of the c.m. energy $\sqrt{s_{NN}}$
    (bottom scale) and beam kinetic energy $T_p$ (upper scale).
    The present inclusive results are shown as full triangles
    and circles, respectively, together with more HADES data
    obtained at bombarding energies of 1.25~\gev\ \cite{hades_nn}
    and 3.5~\gev\ \cite{hades_p35p}.
    Fits to a compilation of $1\pi$ and $1\pi+2\pi$ cross sections \cite{bystricky}
    are shown as solid and long-dashed curves, respectively. 
    Open squares are $\eta$ exclusive cross sections taken
    from \cite{baldini,moskal}; the solid square is the exclusive HADES
    point from \cite{hades_p22p}.  The short-dashed curve corresponds
    to the resonance model \cite{teis}, and the dot-dashed curve
    is the parametrization of inclusive $\eta$ production from \cite{sibirtsev}. }
  \vspace*{-0.2cm}
  \label{cross_inclusive}
\end{figure}

The inclusive $\pi^0$ and $\eta$ production cross sections obtained in the
present analysis can be combined with the result from our
\pp\ runs at 1.25 and 3.5~\gev\ to investigate the excitation function of meson
production in the few-\gev\ regime.  Figure~\ref{cross_inclusive} shows these
cross sections as function of $\sqrt{s_{NN}}$ together with exclusive data.
A wealth of information on exclusive pion production in nucleon-nucleon reactions
has indeed been accumulated over the past 50 years (see \cite{bystricky,moskal}
for reviews and \cite{baldini} for a compilation).  Fits to exclusive
$\pi^0$ production cross sections in one-pion and two-pion channels
from \cite{bystricky} are shown in Fig.~\ref{cross_inclusive}.
These processes are quite well understood in terms of nucleon resonance
excitations within the resonance models \cite{bystricky,cao}.
Data on $\pi^0$ production involving three or more pions in the final state is,
however, still scarce and incomplete, although such processes can be expected
to contribute substantially at beam energies above a few \gev.  Indeed,
from an extrapolation of three-pion production data to $T_p$ = 2.2~\gev\ published
recently \cite{pauly} it is estimated that the contributions of the
$\pi^+\pi^-\pi^0$ and $\pi^0\pi^0\pi^0$ channels could add up to as much as 0.5~mb.
Figure~\ref{cross_inclusive} shows in fact very clearly that
around 2.2~\gev\ bombarding energy inclusive pion production stops to be
fully exhausted by the sum of one- and two-pion channels only.

Turning to $\eta$ production we can do a similar comparison.  Here, data
has again been only available for the exclusive channel.  A compilation of
exclusive $\eta$ production cross sections (from \cite{moskal}, extended
with recent HADES results \cite{hades_p22p})
is depicted in Fig.~\ref{cross_inclusive}, as well as the corresponding
resonance model calculation of Teis \etal\ \cite{teis}.  Assuming that $\eta$
production is mediated solely by the $N^*(1535)$ resonance, the latter gives a
reasonable description of the data.  Just like in the case of pion
production, our inclusive cross sections largely exceed the exclusive ones
in the energy range investigated here.  Because a microscopic description
of multi-particle production is not yet at hand, transport models often
make use of cross section parameterizations based on the Lund string
fragmentation model (LSM) \cite{lund}.  Sibirtsev's parameterization of 
$\eta$ production \cite{sibirtsev}, based on the LSM and shown as dot-dashed
line in Fig.~\ref{cross_inclusive}, turns out to be in reasonable agreement
with our inclusive result, although the intended validity range of the LSM
is in fact at much higher beam energies.

\section{Summary}

To summarize, we have presented data on inclusive \ee\ production in the
reaction \pp\ at 2.2~\gev\ beam kinetic energy.  The measured dielectron
cross sections are in good agreement with the DLS result obtained earlier
at 2.09~\gev.  The employed cocktail of \ee\ sources does not saturate our
data at invariant mass around 0.55~\gevcc.  Furthermore, inclusive
$\pi^0$ and $\eta$ production coss sections were deduced, extending the
world body of meson production data.
Such data, besides having their own interest in the context of medium-energy
hadron reactions, represent valuable input for the analysis and simulation of
proton-nucleus and heavy-ion collisions in the same energy regime.


\acknowledgments {
The collaboration gratefully acknowledges the support by
CNRS/IN2P3 and IPN Orsay (France), by SIP JUC Cracow (Poland) (NN202
286038, NN202198639), by HZDR, Dresden (Germany) (BMBF 06DR9059D), by TU
M\"{u}nchen, Garching (Germany) (MLL M\"{u}nchen, DFG EClust 153,
VH-NG-330, BMBF 06MT9156 TP5, GSI TMKrue 1012), by Goethe-University,
Frankfurt (Germany) (HA216/EMMI, HIC for FAIR (LOEWE), BMBF 06FY9100I, GSI
F$\&$E), by INFN (Italy), by NPI AS CR, Rez (Czech Republic) (MSMT
LC07050, GAASCR IAA100480803), by USC - Santiago  de Compostela (Spain)
(CPAN:CSD2007-00042).
}



\end{document}